\documentstyle[12pt,lathuile]{article}
\def \Bs      { {\mathrm{B}}_{\mathrm{s}} }
\def \Bsbar   { {\overline{{\mathrm{B}}}_{\mathrm{s}}} }
\def \Ds      { {\mathrm{D}}_{\mathrm{s}} }
\def \Dsstar  { {\mathrm{D}}_{\mathrm{s}}^\star }
\def \Dstarz  { {\mathrm{D}}^{\star 0} }
\def \Dsmin   { {\mathrm{D}}_{\mathrm{s}}^- }
\def \Dsplus  { {\mathrm{D}}_{\mathrm{s}}^+ }
\def \Dspm    { {\mathrm{D}}_{\mathrm{s}}^\pm }
\def \Bplus   { {\mathrm{B}}^+ }
\def \Dzero   { {\mathrm{D}}^0 }
\def \b       { {\mathrm{b}} }
\def \bl      { {\mathrm{b}}\to \ell }
\def \D       { {\mathrm{D}} }
\def \B       { {\mathrm{B}} }
\def \Dms     { \Delta m_{\mathrm{s}} }
\def \Npur    { N_{\mathrm{pur}} }
\def \amp     { {\cal{A}} }
\def \Zed     { {\mathrm{Z}} }
\def \aone    { {\mathrm{a}}_1 }
\def \aplus   { {\mathrm{a}}_1^+ }

\def \gevc    { {\mathrm{GeV}}/c }
\def \micr    { \mu{\mathrm{m}} }
\def \ips     { {\mathrm{ps}}^{-1} }

\begin{document}

\title{ \boldmath
REVIEW OF EXPERIMENTAL SEARCHES FOR $\Bs$ OSCILLATIONS  \unboldmath
}

\author{
Duccio Abbaneo \\
{\em CERN, CH-1211, Geneva 23}
}
\maketitle
\baselineskip=14.5pt
\begin{abstract}
The current status of experimental searches for $\Bs$ oscillations
is reviewed. The three ALEPH analyses have been improved since
Summer 2001, and submitted for publication. The combination
of all available analyses is presented and discussed.
\end{abstract}
\baselineskip=17pt
\newpage
\section{Introduction}
\label{sec:intro}

\begin{figure}[bt!]
  \vspace{7.7cm}
  \includegraphics{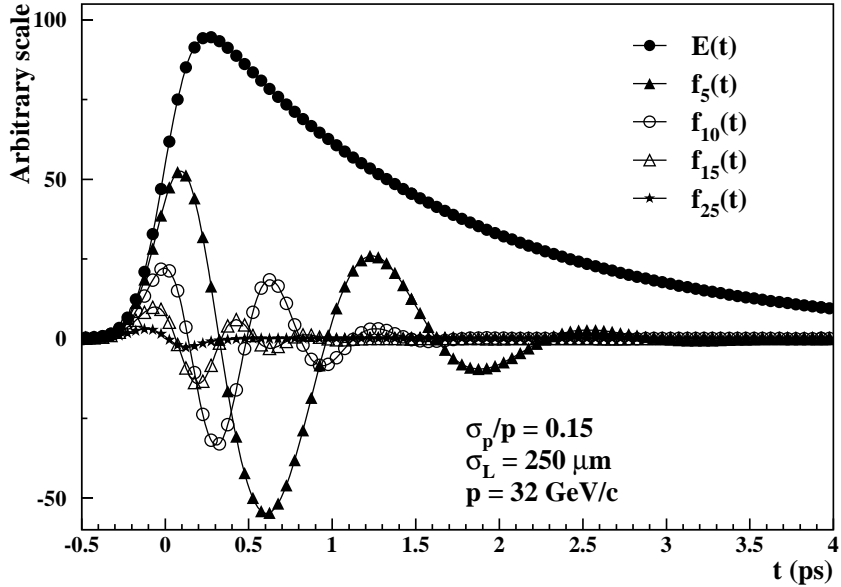}
  \caption{\it
    Effect of proper time  resolution on
the decay exponential E($t$), and on the difference between
the proper time distributions of unmixed and mixed decays,
for several values of the oscillation frequency.
Only monochromatic $\Bs$ mesons are
considered, with perfect separation between mixed and unmixed decays;
resolutions are taken to be Gaussian.
\label{Fig:time} }
\end{figure}

A search for $\Bs$ oscillations consists in detecting the
time-dependent difference between the proper time
distributions of ``mixed'' and ``unmixed'' decays,
\begin{eqnarray}
  {\cal P}(t)_{\Bs \to \Bsbar} & = &
  \frac{\Gamma e^{-\Gamma t}}{2} \, [1 - \cos (\Dms \, t)] \ ,
  \nonumber \\
  {\cal P}(t)_{\Bs \to \Bs} & = &
  \frac{\Gamma e^{-\Gamma t}}{2} \, [1 + \cos (\Dms \, t)] \ .
  \label{eq:pdf}
\end{eqnarray}
The amplitude of such difference is damped not only by the
natural exponential decay,
but also by the effect of the experimental resolution in the proper time
reconstruction. The proper time is derived from the measured decay length
and the reconstructed momentum of the decaying meson. The resolution on the decay length
$\sigma_L$
is to first order independent of the decay length itself, and is largely determined
by the tracking capabilities of the detector. The momentum resolution $\sigma_p$
strongly depends on
the final state chosen for a given analysis, and is typically proportional to the momentum
itself.
The proper time resolution can be therefore written as
\begin{equation}
  \sigma_t = \frac{m}{p} \sigma_L \oplus \frac{\sigma_p}{p} \, t \ ,
  \label{eq:timeres}
\end{equation}
where the decay length resolution contributes a constant term, and the momentum resolution
a term proportional to the proper time. Examples of the observable difference between the 
proper time distributions of mixed and unmixed decays are shown in Fig.~\ref{Fig:time}, 
for the simple case of monochromatic $\Bs$ mesons,
fixed Gaussian resolutions on momentum and decay length (${\sigma_p}/{p}=0.15$,
${\sigma_L}=250\ \micr$),
and for different values of the oscillation frequency.
For low frequency several periods can be observed. As the frequency
increases, the
effect of the finite proper time resolution becomes more relevant,
inducing an overall decrease of observed difference, and a faster damping as a function
of time (due to the momentum resolution component).
In the example given, for a frequency
of $25\ \ips$ only a small effect corresponding to the
first half-period can be seen.

The experimental knowledge of $\Bs$ oscillations available today
comes mostly from SLD and the LEP experiments. At SLD the
lower statistics (less than $4\times 10^5$ hadronic $\Zed$ decays,
compared to almost $4\times 10^6$ for each LEP experiment)
is compensated, especially at high frequency,
by the excellent tracking performance provided by the small
and precise CCD vertex detector: at SLD in a typical analysis
a decay length resolution of $70-80\ \micr$ is obtained for
a core of $60\%$ of the events, with tails of $250-300\ \micr$,
while at LEP the corresponding figures are $200-250\ \micr$ and
$700-1000\ \micr$.

\section{Analysis methods}

Analyses based on different final state selections have been developed
over the years, which offer different advantages in terms of
statistics, signal purity and resolution.
The selection criterion chosen also determines
the strategy for tagging the flavour of the $\Bs$ meson
at decay time. The flavour at production time is tagged
from the hemisphere opposite to the one containing the selected
candidate, as well as 
from the fragmentation particles belonging to the candidate
hemisphere.
Finally, the proper time is reconstructed and the oscillation
is studied by means of a likelihood fit to the proper
time distributions of mixed and unmixed decays.

\subsection{Selection methods and flavour tagging at decay time}

\begin{figure}[bt!]
  \vspace{5.2cm}
  \includegraphics{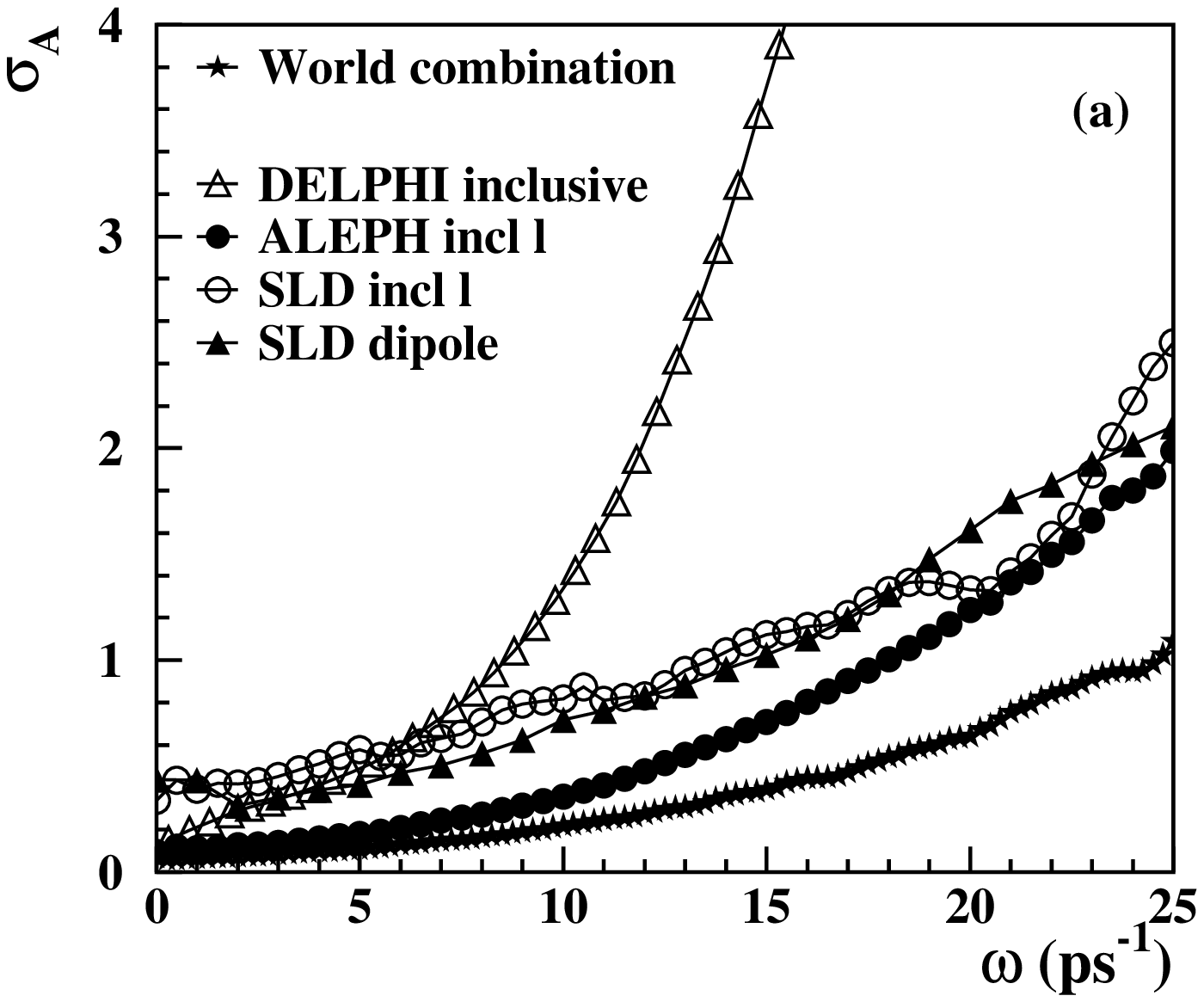}
  \includegraphics{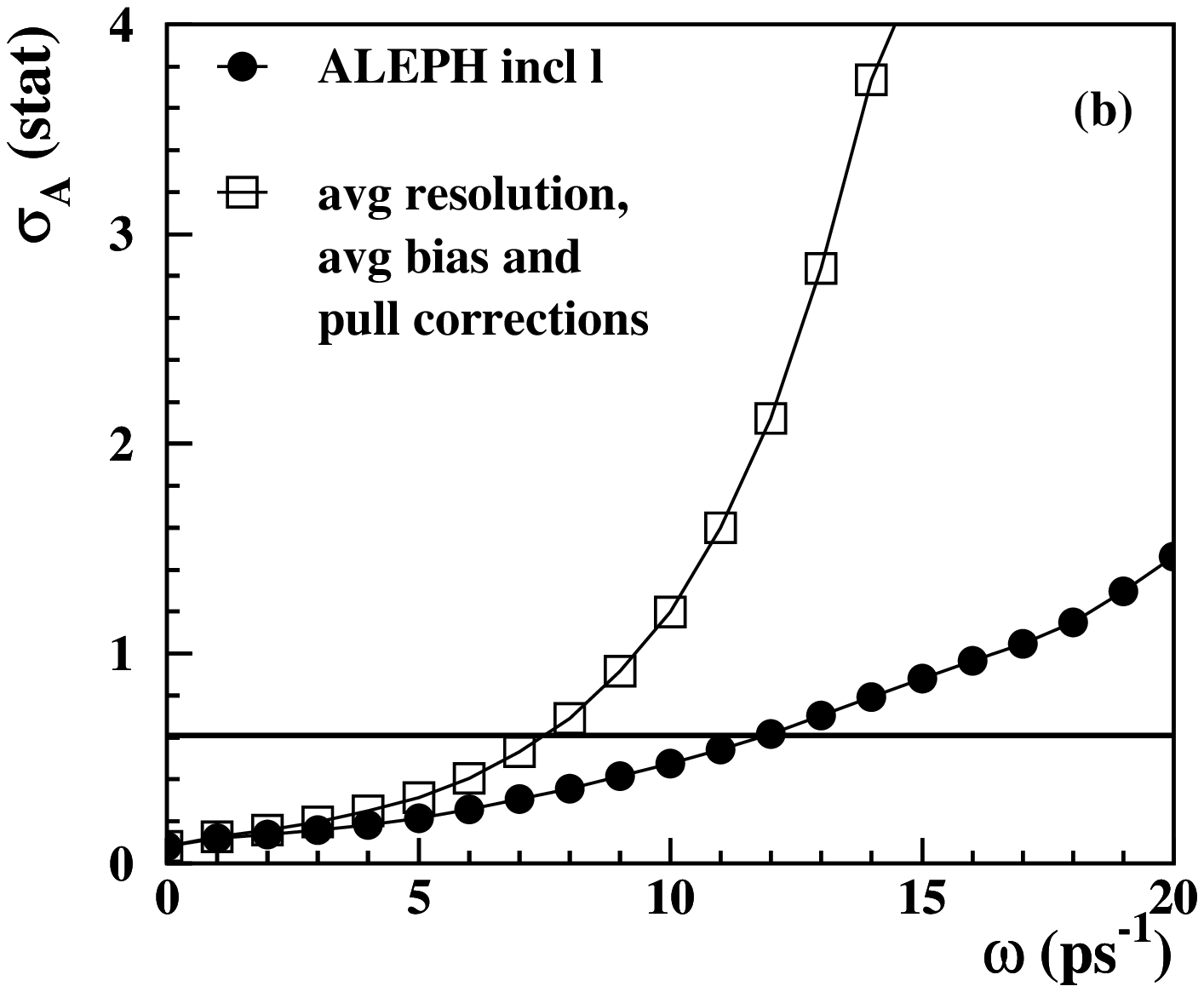}
  \caption{\it (a) The SLD analyses based on fully inclusive and inclusive lepton selections 
are compared to the corresponding DELPHI and ALEPH analyses. The error on the
measured oscillation amplitude is shown as a function of test frequency. \newline
(b) Error on the measured oscillation amplitude for the ALEPH inclusive lepton analysis,
compared to the same analysis when average values are used for the momentum and
decay length resolutions (see text).
\label{Fig:compsld} }
\end{figure}

\indent In {\em fully inclusive selections} no decay product of the
$\Bs$ is explicitly identified, which implies that no
straightforward method is available to tag the flavour at decay
time. Variables sensitive to the sign of the
 charge flow between the secondary and the tertiary vertices
 ($\Bs \to
\Dsmin$, $\Bsbar \to \Dsplus$) can be built to discriminate
 between $\Bs$ and $\Bsbar$ decays. This method  gives
the highest statistics (a few $10^5$ candidates at LEP, about
$10^4$ at SLD); the signal fraction is what is given by nature
($\approx 10\%$). The reconstruction of the secondary vertex is
completely based on topology, which requires excellent tracking
capabilities, to avoid that the track mis-assignment spoils the
decay length reconstruction. The method has high performance at
SLD; at LEP it has been attempted by DELPHI: the much more
difficult experimental environment results in uncertainties on the
fitted amplitude (Section~\ref{sec:ampl}) raising fast with
increasing frequency (Fig.~\ref{Fig:compsld}a).

In {\em semi-inclusive selections} at least one $\Bs$
decay product, typically a lepton, is explicitly identified.
The charge of such a particle provides the final state tag.
Statistics are still high (a few $10^4$ at LEP, a few $10^3$ at SLD),
the signal purity is typically still about $10\%$, or slightly
enhanced if more than one decay product is identified.
This method gives excellent performance both at SLD and at LEP.
The ALEPH inclusive lepton analysis is the most powerful
available to date (Fig.~\ref{Fig:compsld}a).

In {\em semi-exclusive selections}  the $\Ds$ meson from the $\Bs$ meson decay
is fully reconstructed; it is sometimes
combined with a tagged lepton,
in which case only the neutrino
is undetected. These methods have substantially lower statistics, but the
signal purity can be as high as $60\%$, giving interesting performance
both at LEP and SLD.

{\em Fully exclusive selections} have been used at LEP by ALEPH
and DELPHI, obtaining about 50 candidates only,
reconstructed in many different decay channels, with $\approx 50\%$
average signal purity. Nevertheless, since all the decay products
are identified, these events have excellent proper time
resolution, and therefore give useful information in the high
frequency range. The little statistics make this method not
suitable for SLD.

\subsection{Flavour tagging at production time}

All particles of the event except those tagged as the $\Bs$ meson
decay products can be used to derive information on the $\Bs$ meson
flavour at production time.

In the hemisphere containing the $\Bs$ meson candidate, charged particles
originating from the primary vertex, produced in the hadronization of the
b quark, may retain some memory of its charge. Either the charges of all
tracks are combined, weighted according to the track kinematics, or the track
closer in phase space to the  $\Bs$ meson candidate is selected, requiring that it be
compatible with a kaon.

In the opposite hemisphere, the charge of the other b hadron can be tagged, exploiting
the fact that b quarks are produced in pairs. Tracks from the b-hadron decay
can be distinguished from fragmentation tracks on a statistical basis, from their
compatibility with the primary vertex or with an inclusively-reconstructed secondary
vertex. Inclusive hemisphere-charges can be formed assigning
weights to  tracks according to the
probability that they belong to the primary or the secondary vertex, complemented with
more ``traditional'' jet-charges, where the weights are defined on the basis of track
kinematics.
Furthermore, specific decay products, such as leptons or kaons,
can be searched for also in the opposite hemisphere, and their charge used
as an estimator of the quark charge.

Finally, both at LEP and at SLD, the polar angle of the $\Bs$
meson candidate is also correlated with the charge of the quark,
because of the forward-backward asymmetry in the $\Zed$ decays.
This correlation is particularly relevant at SLD, due to the
polarization of the electron beam.

Many of the production-flavour estimators mentioned above are correlated
among themselves. The most recent and sophisticated analyses have attempted to
use efficiently all the available information by combining the different estimators
using neural network techniques.

\subsection{The proper time reconstruction and the oscillation fit}
\label{Sec.ptim}

As discussed above in Section~\ref{sec:intro}, the proper time of the decaying meson 
is reconstructed from the measured decay length and the estimated momentum. 
The uncertainties from the fit to the primary and secondary vertex positions are used to estimate, 
event by event, the uncertainty on the reconstructed decay length: such estimate 
needs in general to be inflated ({\em pull correction})
to account for missing or mis-assigned particles 
(except, possibly, for the case of fully reconstructed candidates). 
Pull corrections, possible bias corrections, as well as the estimated 
uncertainty on the reconstructed momentum, must be parameterized as a function of 
the relevant variables describing the event topology, in order to give to each event 
the appropriate weight in the analysis.
Such a procedure is crucial to achieve good performance at high frequency, especially 
for inclusive analyses, that deal with a wide variety of topologies.
As an example, in Fig.~\ref{Fig:compsld}b the errors on the measured amplitude as a 
function of test frequency is given for the ALEPH inclusive lepton analysis, 
compared to what is obtained using average figures for the decay length 
and momentum resolutions: the performance at high frequency is completely spoiled in this case.

In the most sophisticated analyses also the sample composition and the initial and final state mis-tag probabilities are evaluated as a function of discriminating variables, and used event by event in the fit, so that in fact each event is treated as a single experiment with its own signal purity, flavour tag performance and proper time resolution.

\section{The amplitude method}
\label{sec:ampl}

The analyses completed so far are not able to resolve 
the fast $\Bs$ oscillations: they can only exclude a 
certain range of frequencies. Combining such excluded ranges
is not straightforward, and a specific method, called ``amplitude method''
has been introduced for this purpose\cite{hgm}. In the likelihood fit to 
the proper time distribution of decays tagged as mixed or unmixed, the 
frequency of the oscillation is not taken to be the free parameter,
but it is instead fixed to a ``test'' value $\omega$. An auxiliary 
parameter, the  amplitude $\amp$ 
of the oscillating term is introduced, and left free in the fit.
The proper time distributions for unmixed and unmixed decay, prior to 
convolution with the experimental resolution, are therefore written as
\begin{equation}
  {\cal P}(t) = 
  \frac{\Gamma e^{-\Gamma t}}{2} \, [1 \pm \amp \cos (\omega\, t)] \ ,
\end{equation}
with $\omega$ the test frequency and $\amp$ the only free parameter. 
When the test frequency is much smaller than the true frequency 
(\mbox{$\omega \ll \Dms$}) the expected value for the amplitude is 
\mbox{$\amp = 0$}.
At the true frequency (\mbox{$\omega = \Dms$}) the expectation is 
\mbox{$\amp = 1$}.
All values of the test frequency $\omega$ for which 
\mbox{$\amp + 1.645 \sigma_\amp < 1$} are excluded at $95\%$~C.L.

The amplitude has well-behaved errors, and 
different measurements can be combined in a straightforward way, 
by averaging the amplitude measured at different test frequencies.
The excluded range is derived from the combined amplitude scan.

\begin{figure}[tb!]
 \vspace{4.9cm}
  \includegraphics{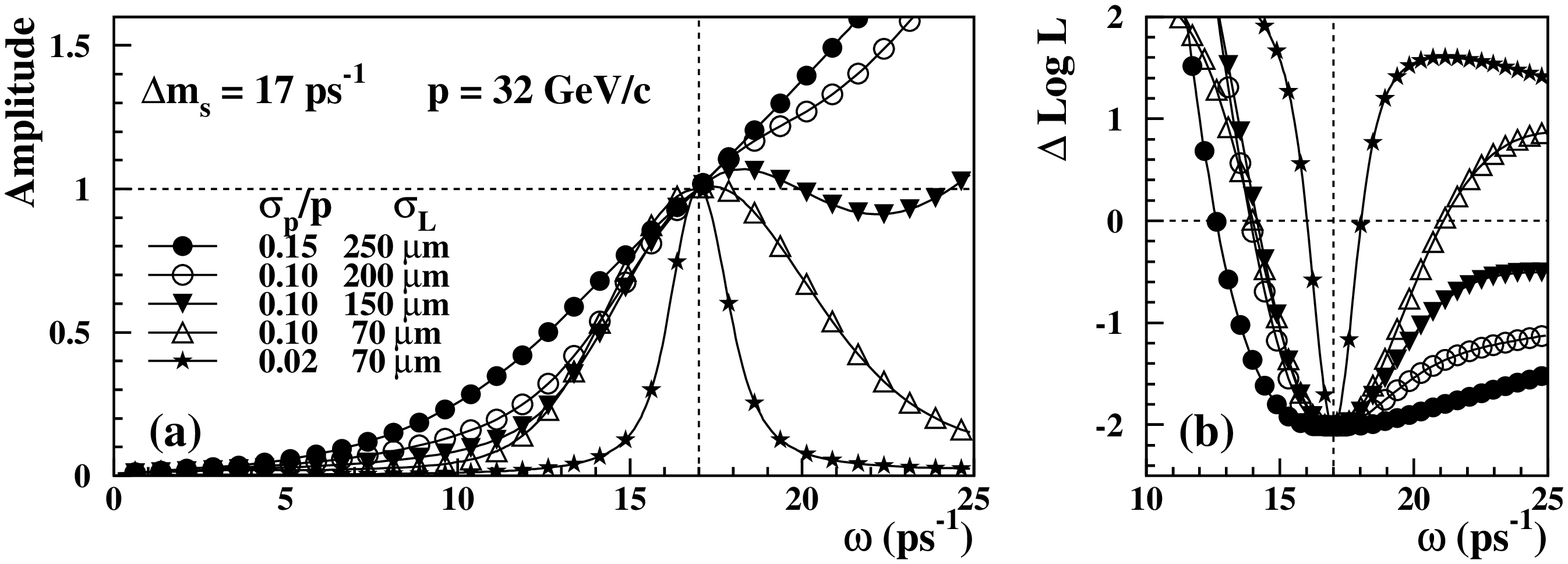}
 \caption{\it
(a) Expected amplitude shape for a true frequency $\Dms = 17 \ \ips$, monochromatic $\Bs$ mesons of $p=32\ \gevc$ and different values of momentum and decay length resolutions (taken to be Gaussian) \newline
(b) Corresponding expected shape of the likelihood profile. The vertical scale 
for each curve is arbitrary as it depends on the statistics available for the analysis.
    \label{Fig:ampl} }
\end{figure}

An interesting issue is what should be expected for the shape of the measured amplitude 
in the vicinity of the true frequency $\Dms$, and beyond it. 
The shape can be calculated analytically\cite{noi} for simple cases: 
some examples are shown in Fig.~\ref{Fig:ampl}a. The expected shape varies 
largely depending on the decay length and momentum resolution: the only solid 
features are that the expectation is $\amp=1$ at the true frequency and 
$\amp=0$ far below it. The amplitude spectrum can be converted back into 
a likelihood profile as a function of frequency\cite{noi}, that is always 
expected to show a minimum at the value of the true oscillation frequency, 
as shown in Fig.~\ref{Fig:ampl}b.

\section{The three ALEPH analyses}

ALEPH has recently submitted for publication three analyses, based on fully reconstructed candidates, 
$\Ds \ell$ final states, and inclusive semileptonic
decays, respectively.

\begin{figure}[t!]
  \vspace{6.8cm}
  \includegraphics{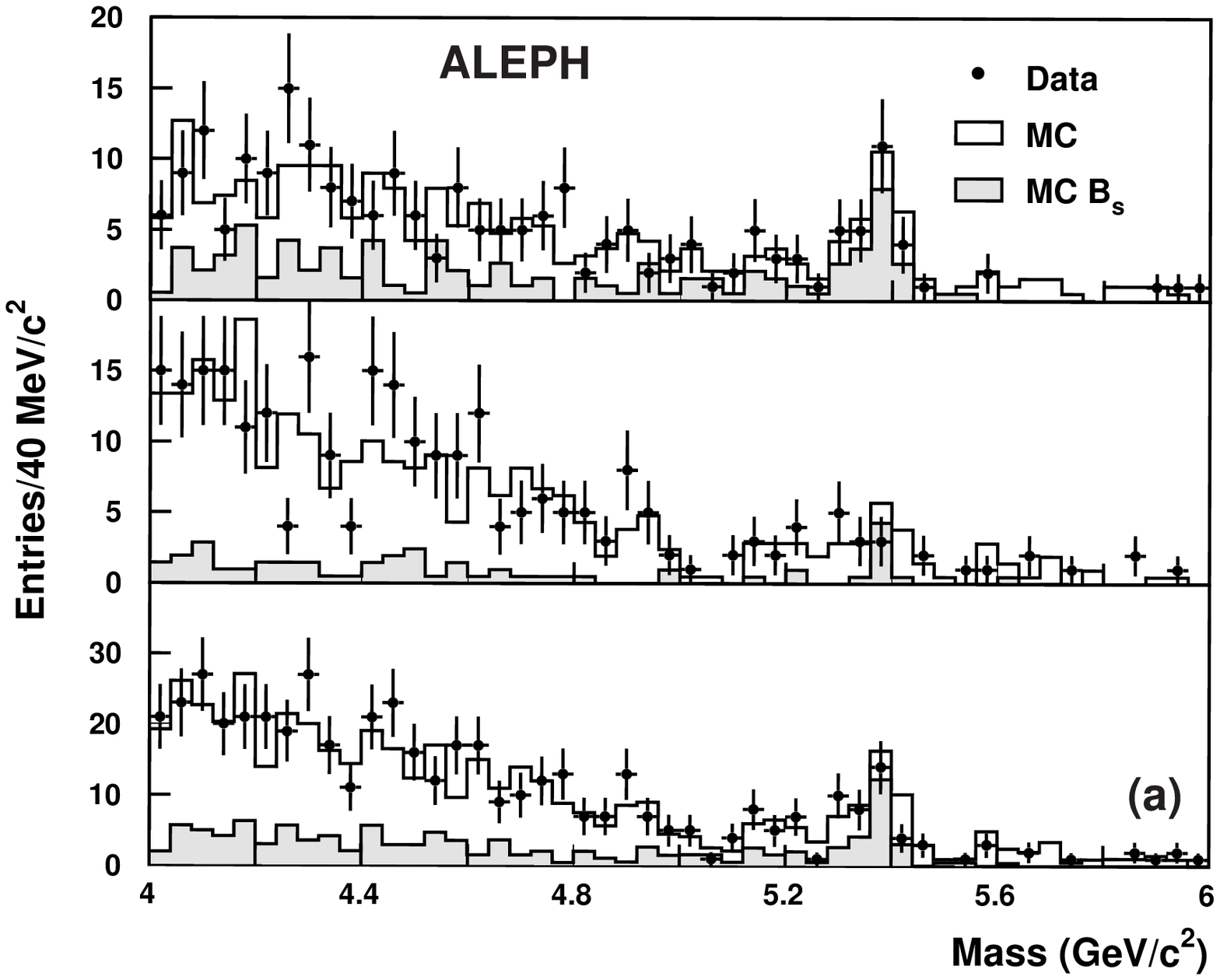}
  \includegraphics{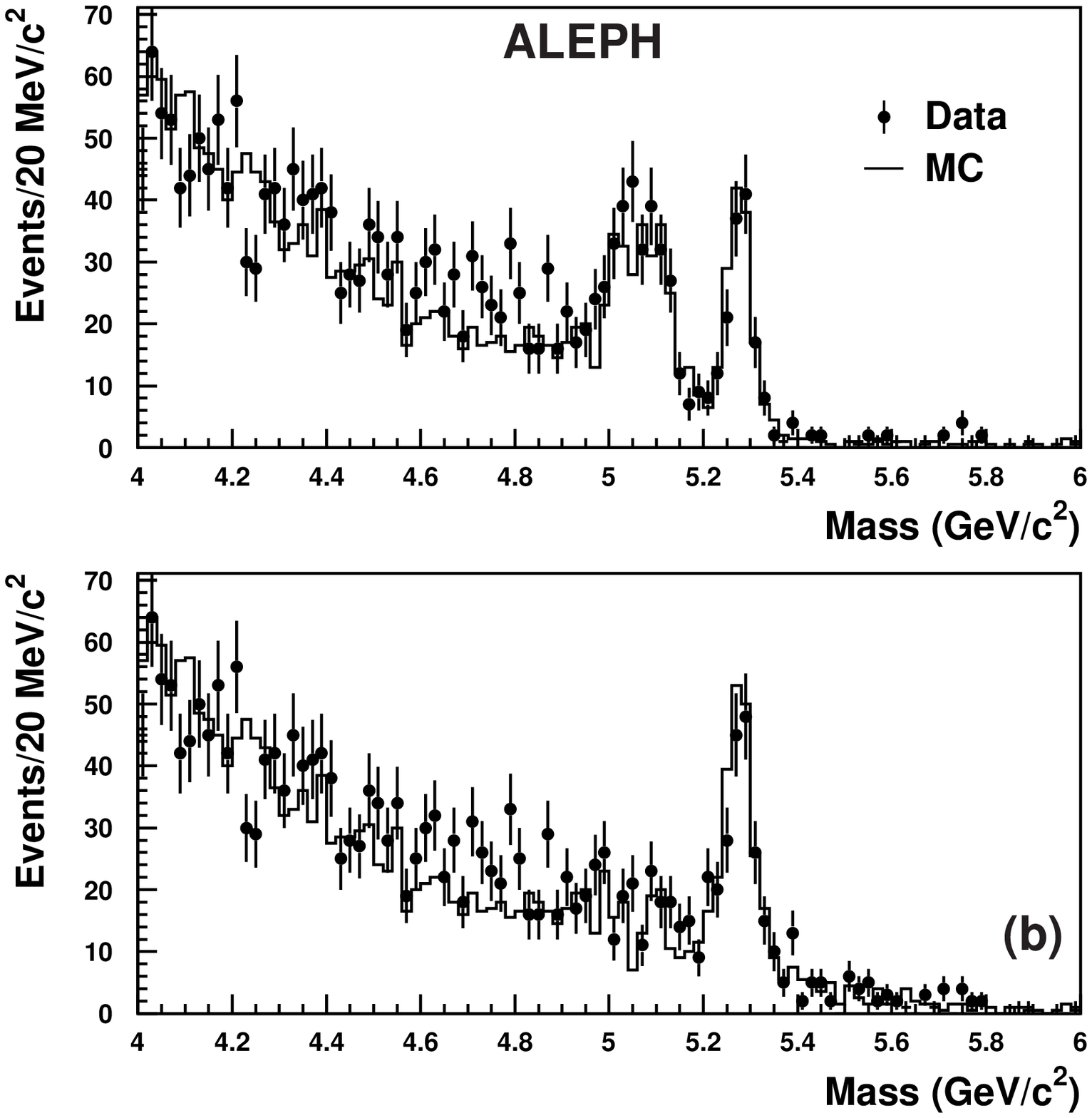}
   \caption{\it
(a) Invariant mass distribution for candidates reconstructed in the  $\Dsmin \pi^+$ final state (top), 
$\Dsmin \aplus$ 
final state (middle) and all channels together (bottom). \newline
(b) Invariant mass distribution of $\Bplus$ candidates before (top) and after (bottom) $\pi^0$ reconstruction.
\label{Fig:bsmass} }
\end{figure}

Hadronic $\Bs$ decays are reconstructed in the decay modes $\Dsmin \pi^+$ and $\Dsmin \aplus$ 
(charge conjugate modes are implied), with the $\Ds$ and the $\aone$ decaying to charged particles only. 
Other modes, with a $\Dsstar$ decaying to $\Ds \gamma$ or a $\rho^+$ decaying o $\pi^+ \pi^0$, are 
also fully reconstructed, by looking for photons kinematically compatible with being $\Bs$ decay products. 
The mass region below the $\Bs$ peak is used in the analysis, as weel as the main peak,
as it presents an interesting $\Bs$ purity, 
due to decays involving a $\Dsstar$ or a $\rho$, where the photon or the $\pi^0$ were not reconstructed.
The distribution of the invariant mass for the reconstructed candidates is shown in Fig.~\ref{Fig:bsmass}a. 
The accuracy of the Monte Carlo simulation in reproducing the reconstruction efficiency, especially for what concerns 
photon and $\pi^0$ reconstruction, is checked using $\Bplus$ decays to $\Dzero \pi^+$. In this case, 
decays to final states with a $\Dstarz$ decaying to $\Dzero \pi^0$ give the same number and type of charged particles, 
and are reconstructed below the $\Bplus$ mass peak. The agreement between data and Monte Carlo, before and after
the reconstruction of the additional $\pi^0$ is excellent, as shown in Fig.~\ref{Fig:bsmass}b.

\begin{figure}[b!]
  \vspace{6.2cm}
  \includegraphics{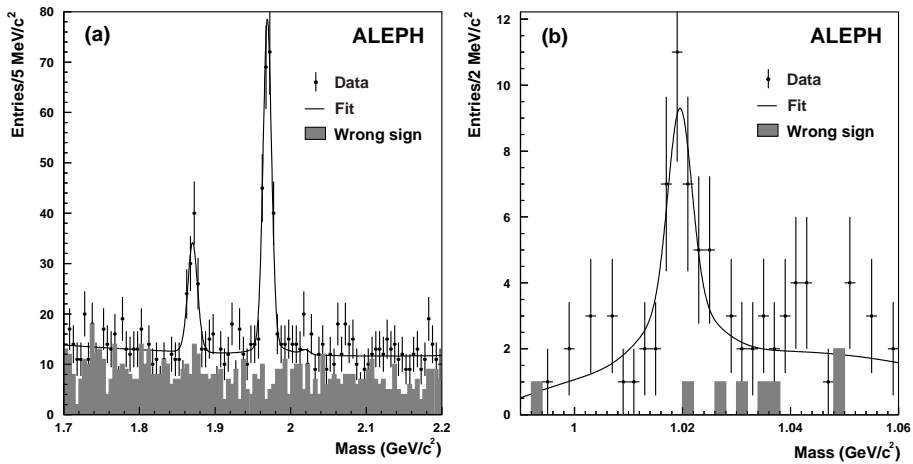}
  \caption{\it
(a) Invariant mass distribution of $\Ds$ candidates in $\Dsmin \ell^+$ combinations, reconstructed in
hadronic final states. \newline
(b) Mass distribution of the selected $\phi$ candidates in semileptonic $\Ds$ decays.
    \label{Fig:dsmass} }
\end{figure}

Final states containing a $\Ds \ell$ pair are reconstructed using several hadronic modes of the $\Ds$, plus the
decay mode $\Dsmin \to \phi \ell^-$. The fraction of resonant component is estimated directly from the data, 
from the sidebands of the $\Ds$ and $\phi$ mass peaks, shown in Fig.~\ref{Fig:dsmass}. A resonant background 
component is given by double-charm b decays $\b \to \Dspm \D (X)$ with $\D \to X \ell^\mp$. Such events are 
discriminated from the signal using lepton kinematics and jet topology: several variables are combined by means 
of a neural network, and the signal purity is estimated as a function of the discriminant, using simulated events.

\begin{figure}[bt!]
  \vspace{12cm}
  \includegraphics{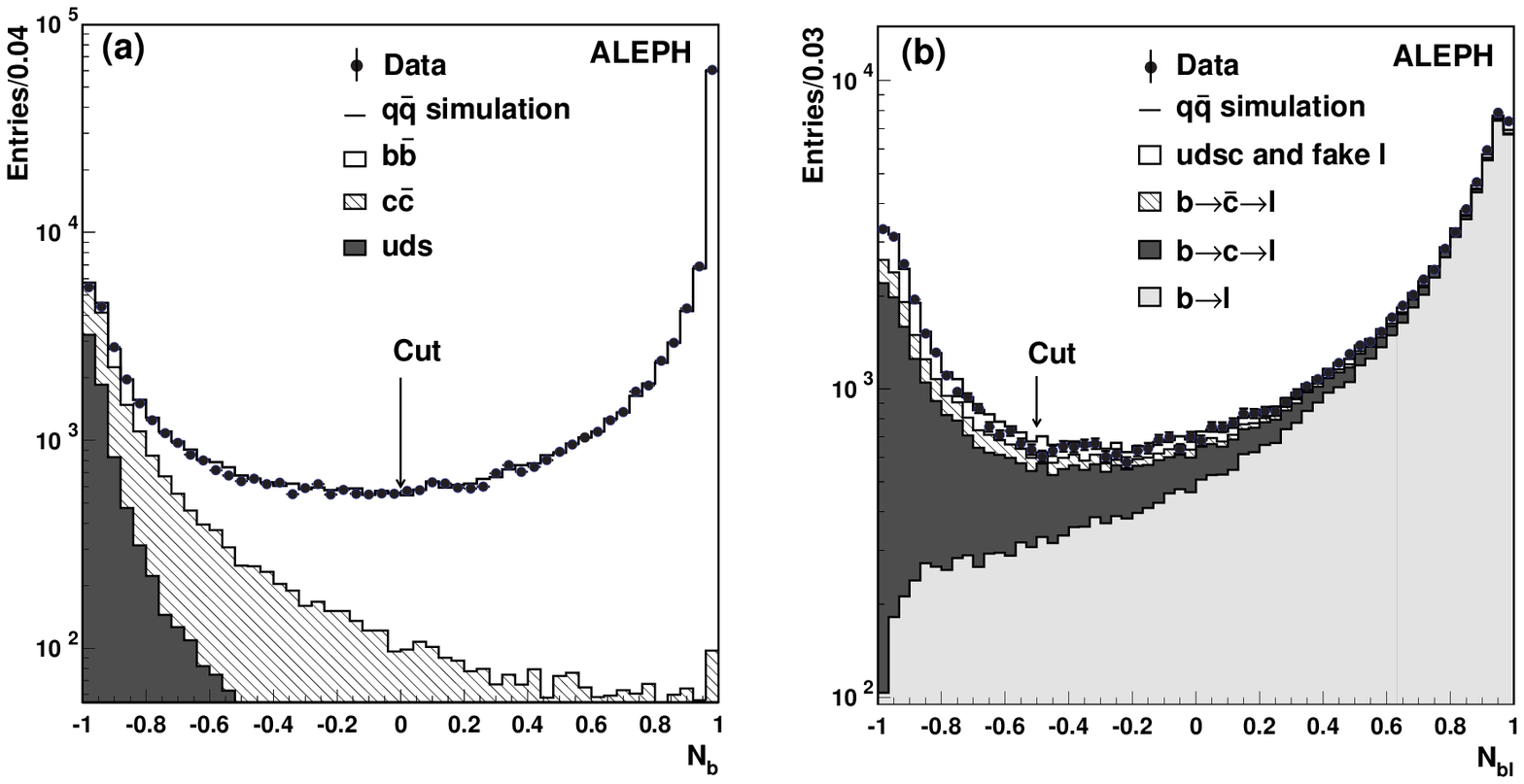}
  \includegraphics{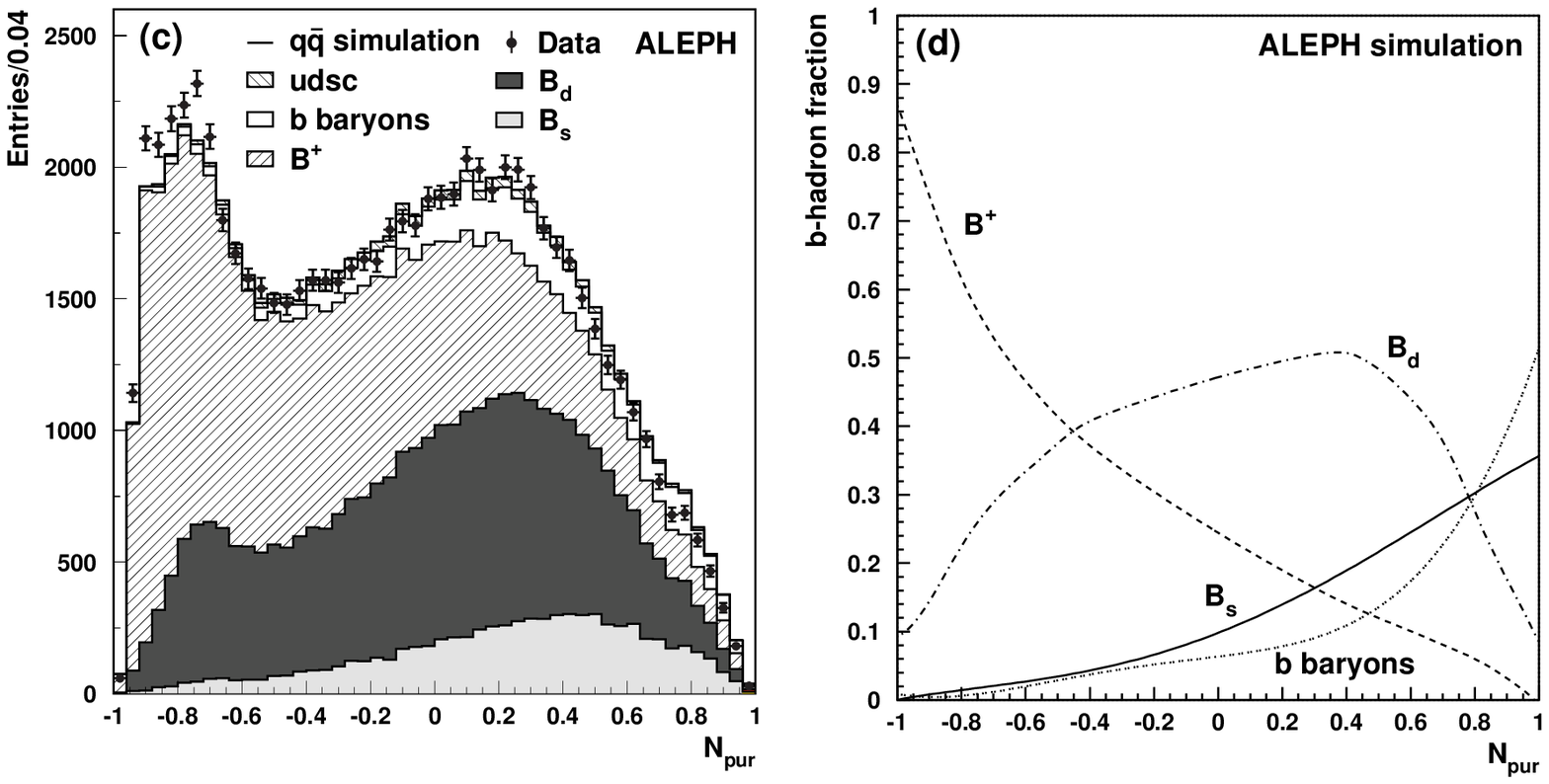}
  \caption{\it
    The b-tagging variable (a) and the $\bl$ tagging variable (b) distributions for the inclusive
lepton candidates. Distribution of the $\Bs$ purity variable $\Npur$ (c), and fraction of each
b-hadron species as a function of $\Npur$ (d).
    \label{Fig:incl_distr} }
\end{figure}

The analysis of inclusive semileptonic decays makes use of a novel algorithm for secondary vertex
reconstruction. A $\D$ track is defined from the charged particles identified as decay products of
the inclusively reconstructed $\D$ meson. The direction of such $\D$ track is improved by adding
photons reconstructed in a cone around the $\D$ meson, and forming with it an invariant mass smaller
than the nominal $\D$ mass. A $\B$ track is formed from the jet or the thrust axis direction, 
depending on the event topology, with angular uncertainties parameterized from the simulation.
The lepton, the improved $\D$ track and the $\B$ track are fit together to find the $\B$ meson 
decay vertex. 
As mentioned in Section~\ref{Sec.ptim}, estimates of the initial and final state tag performance, 
of the sample composition (Fig.~\ref{Fig:incl_distr}), and, most importantly, of the
uncertainty on the reconstructed decay length and momentum, are obtained for each event from
parametrizations as a function of the relevant event properties.

\begin{figure}[bt!]
  \vspace{5cm}
  \includegraphics{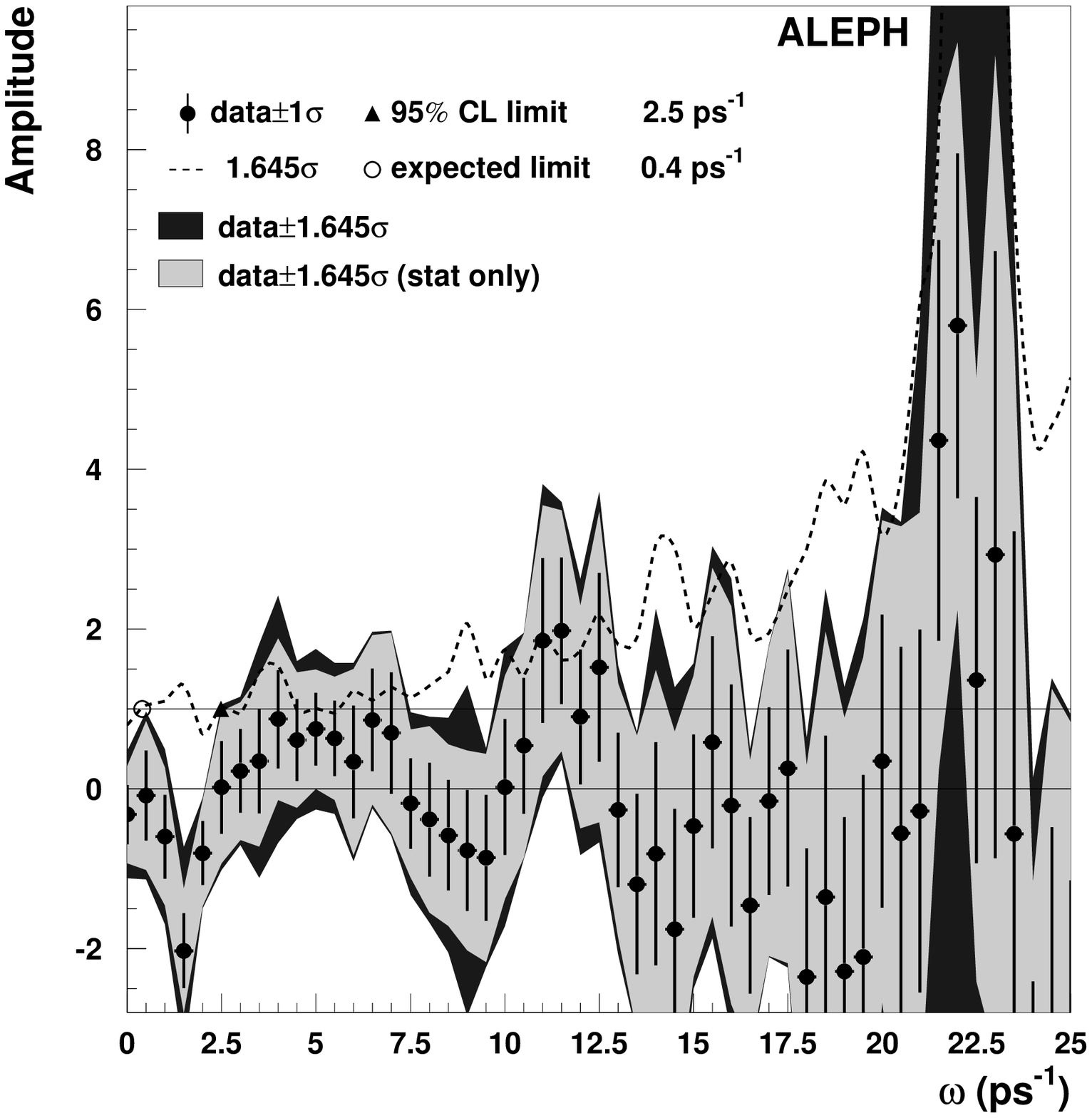}
  \includegraphics{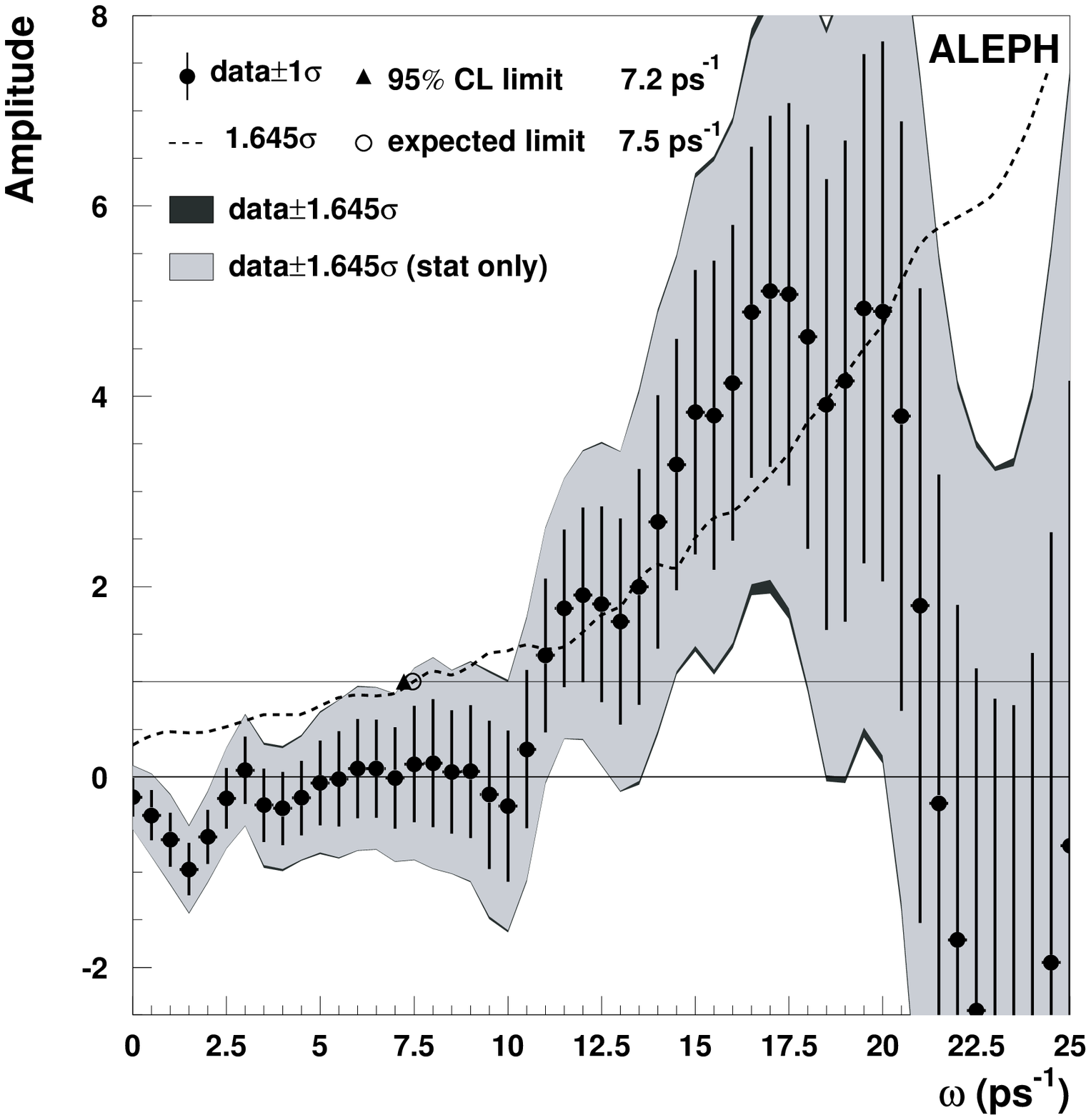}
  \includegraphics{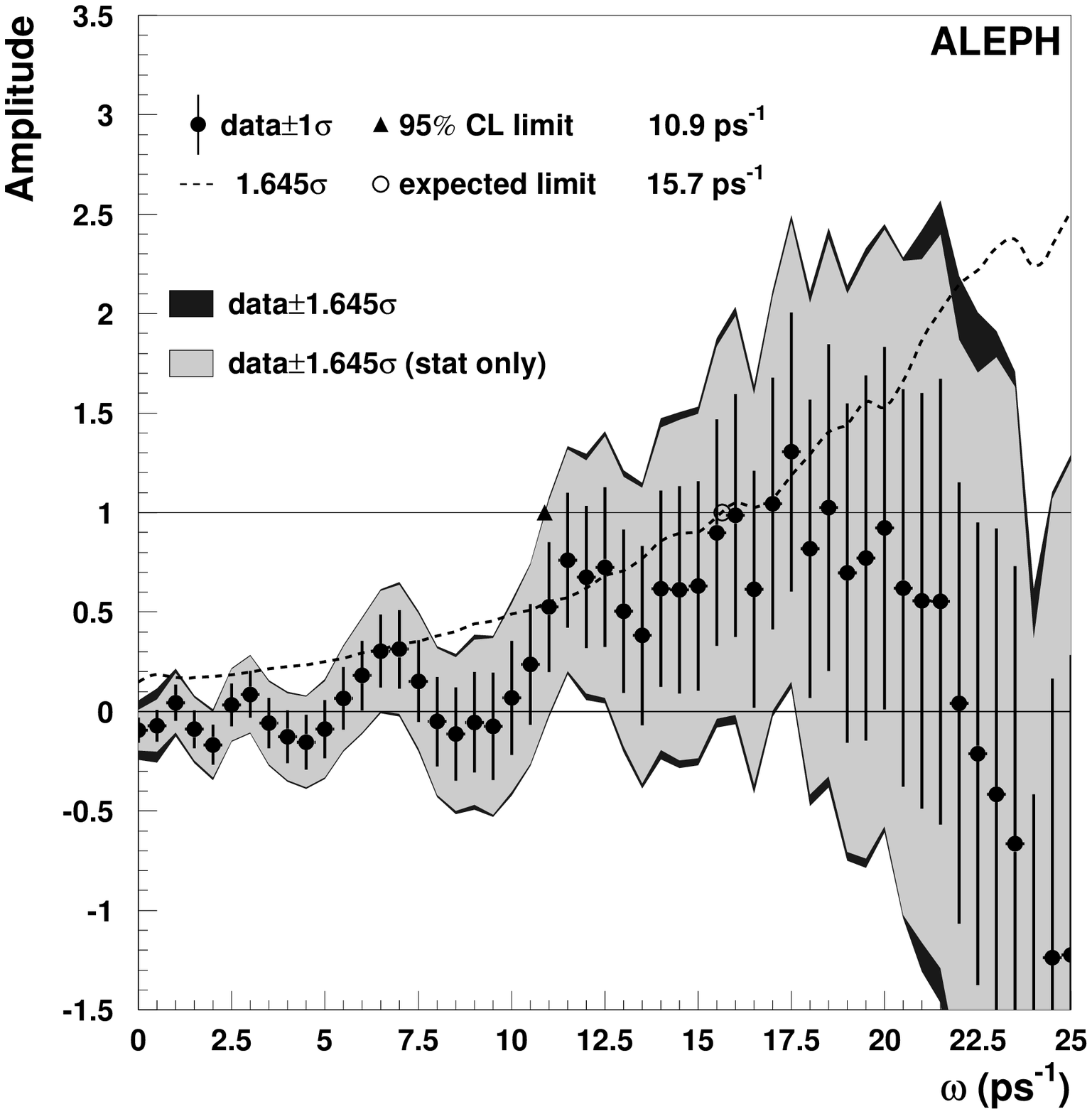}
\vspace{-0.9cm} \hspace{1cm} \scriptsize{\bf (a)} \hspace{4.3cm} \scriptsize{\bf (b)} \hspace{4.3cm} \scriptsize{\bf (c)}
\vspace{0.5cm}
   \caption{\it
    Fitted amplitude spectra for the fully reconstructed sample (a), the $\Ds \ell$ sample (b)
and the inclusive lepton sample (c).
\label{Fig:aleph} }
\end{figure}

For the three analyses, systematic uncertainties are calculated taking into account not only
the change in the fitted amplitude, but also the change in its statistical uncertainty,
which in some cases is more relevant. The method is based on toy experiments.
The amplitude spectra obtained are shown in Fig.~\ref{Fig:aleph}.

\section {Conclusions: the World combination}

\begin{figure}[t!]
  \vspace{8.9cm}
  \includegraphics{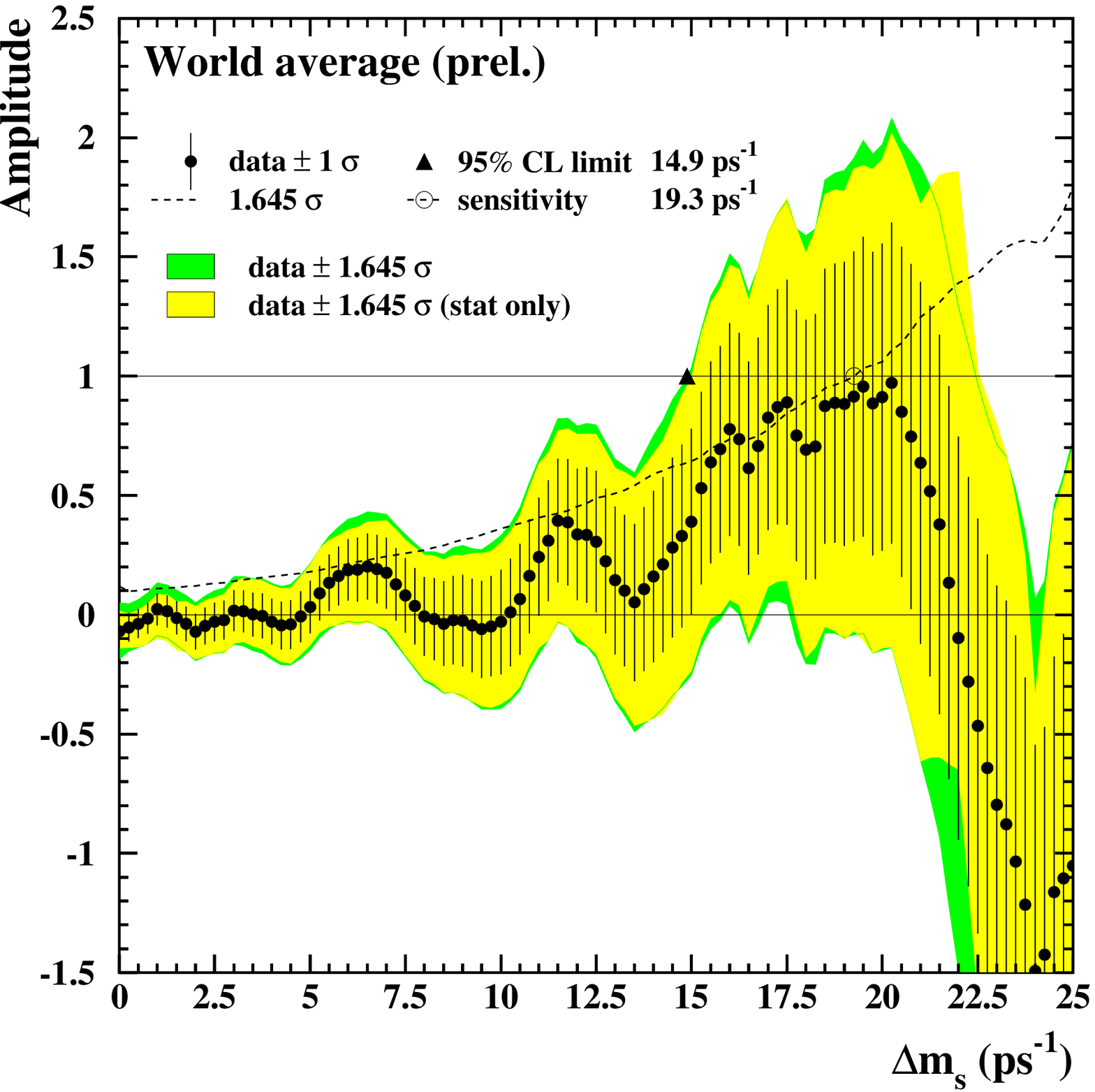}
  \includegraphics{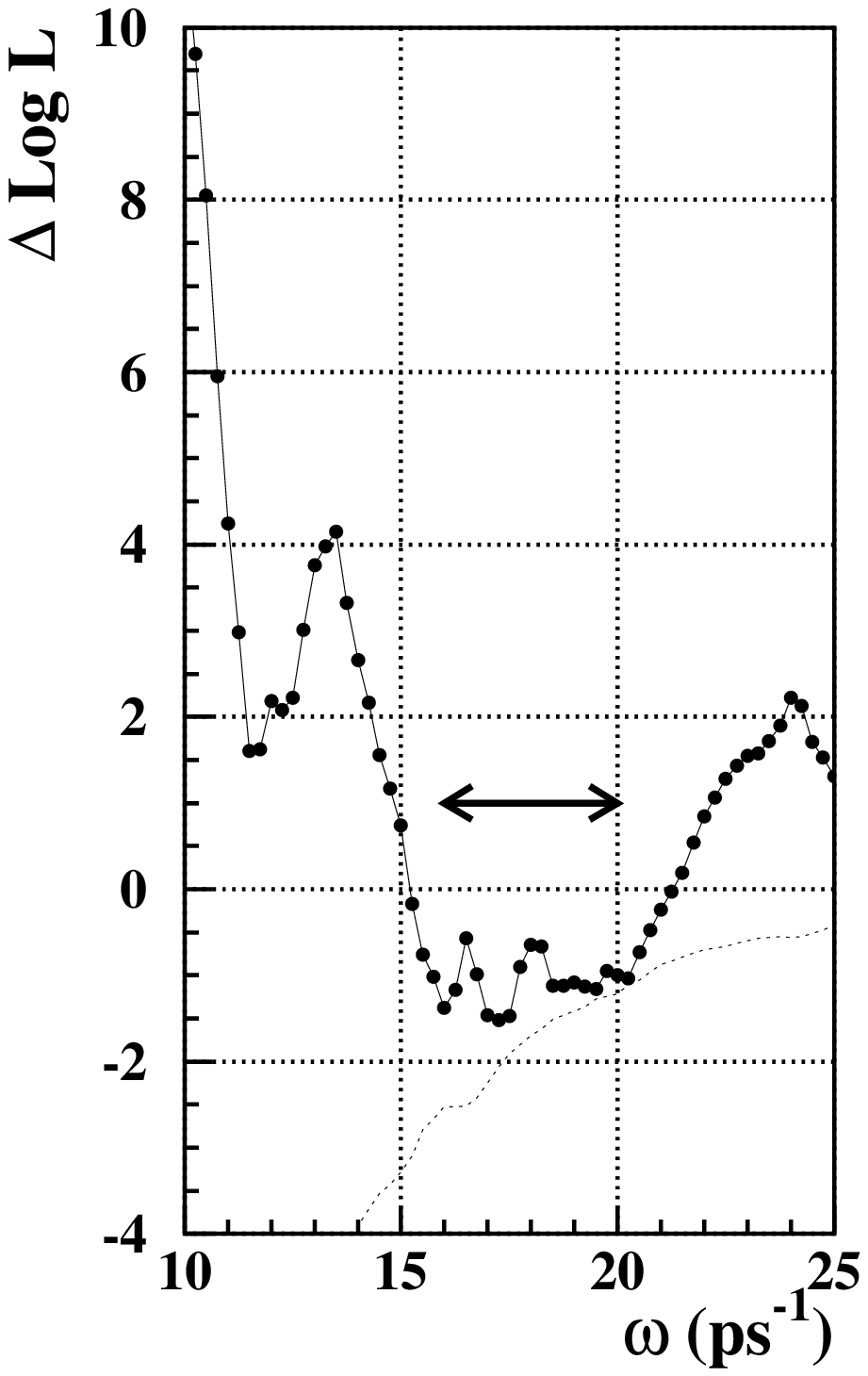}
\vspace{-1.6cm} \hspace{1cm} {\bf (a)} \hspace{9.1cm} {\bf (b)}
\vspace{1cm}
   \caption{\it
Amplitude spectrum for the combination of all available analyses (a), and corresponding
log-likelihood profile (b); the dotted curve gives the expected likelihood depth 
at each frequency for $\omega = \Dms$. \newline
\label{Fig:world} }
\end{figure}

The combined amplitude spectrum from all available 
analyses~\footnote{Up-to-date references for all the available
published and preliminary analyses can be obtained from
http://lepbosc.web.cern.ch/LEPBOSC/references/} is shown in Fig.~\ref{Fig:world}a.
Frequencies smaller than $14.9~\ips$ are excluded at 95\%~CL. The expected
limit, $19.4~\ips$, is substantially higher because  amplitude values
different from zero, and close to unity,  are found in the  frequency range $16 - 20~\ips$.
The structure observed is compatible with the hybothesis of an oscillation signal,
as can be appreciated from the likelihood profile shown in Fig.~\ref{Fig:world}b, but its statistical
significance is smaller than two standard deviations, and is therefore insufficient to claim 
an observation.

\begin{figure}[b!]
  \vspace{7cm}
  \includegraphics{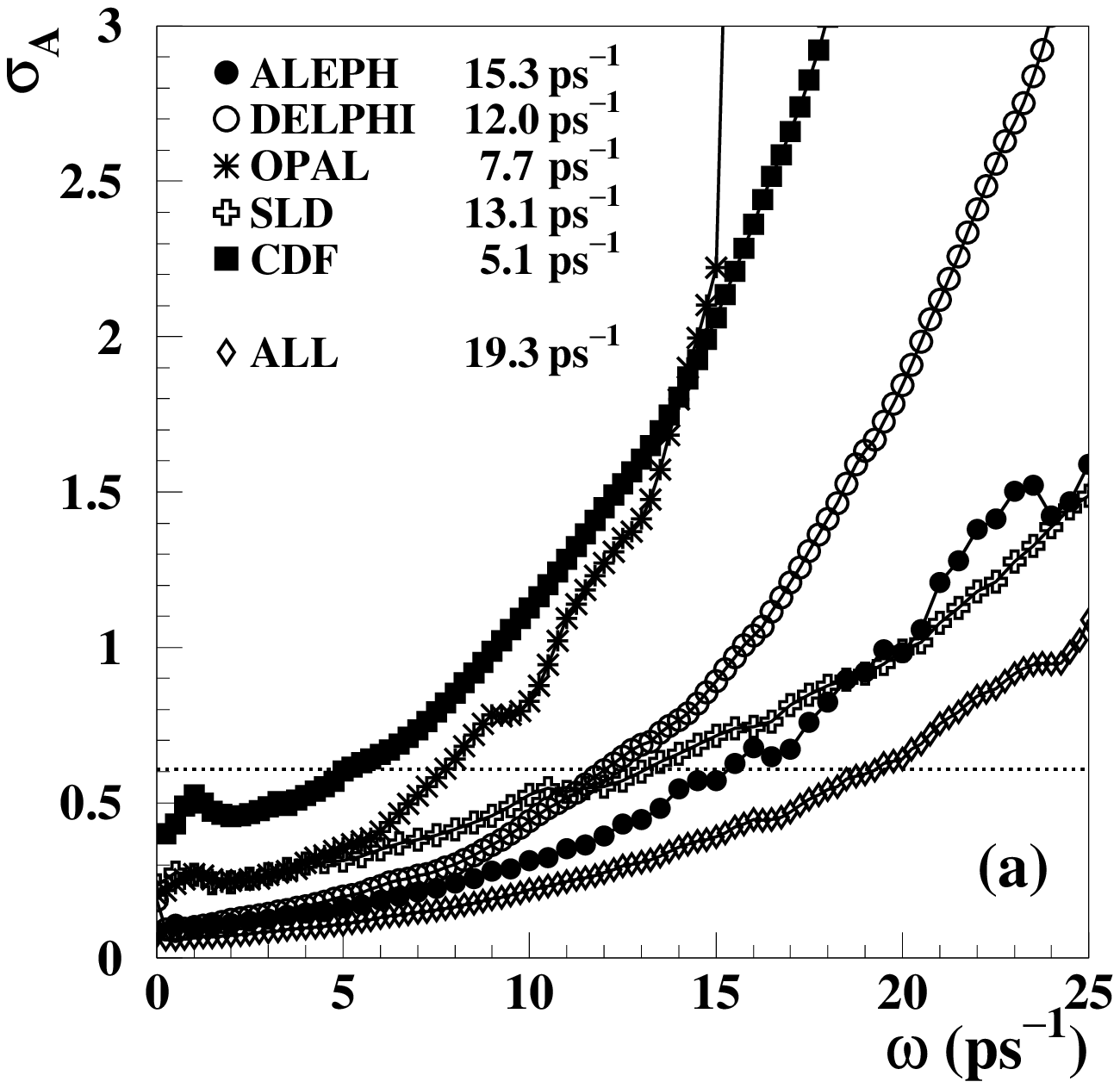}
  \includegraphics{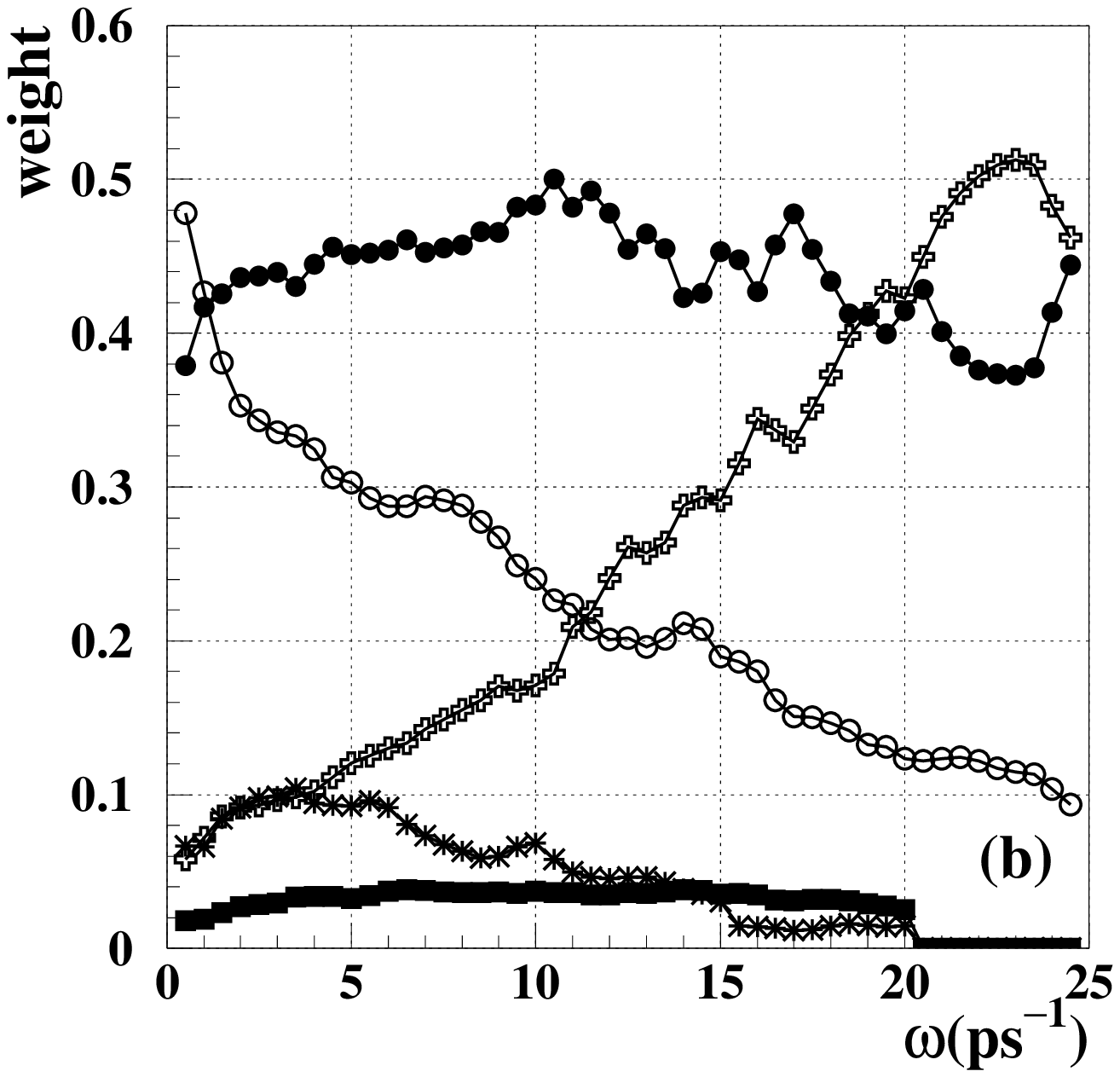}
   \caption{\it
(a) Uncertainty on the fitted amplitude as a function of test frequency, for the experiments
that have performed $\Bs$ oscillation searches. \newline
(b) Weight of the different experiments in the combination.
\label{Fig:compexp} }
\end{figure}

The estimated uncertainty on the fitted amplitude as a function of test frequency is shown in
Fig.~\ref{Fig:compexp}a for the available analyses combined by experiment. In 
Fig.~\ref{Fig:compexp}b the weight of each experiment in the world combination is shown.
In the region $16-20 \ips$ where the deviation from $\amp=0$ is observed, ALEPH and SLD
dominate the combination.

\section{Acknowledgements}
I wish to thank the organizers for the interesting conference
and the pleasant stay in La Thuile.

\end{document}